# Covariate Balance in Simple, Stratified and Clustered Comparative Studies

**Ben B. Hansen and Jake Bowers**

*Abstract.* In randomized experiments, treatment and control groups should be roughly the same—balanced—in their distributions of pretreatment variables. But how nearly so? Can descriptive comparisons meaningfully be paired with significance tests? If so, should there be several such tests, one for each pretreatment variable, or should there be a single, omnibus test? Could such a test be engineered to give easily computed $p$-values that are reliable in samples of moderate size, or would simulation be needed for reliable calibration? What new concerns are introduced by random assignment of clusters? Which tests of balance would be optimal?

To address these questions, Fisher's randomization inference is applied to the question of balance. Its application suggests the reversal of published conclusions about two studies, one clinical and the other a field experiment in political participation.

*Key words and phrases:* Cluster, contiguity, community intervention, group randomization, randomization inference, subclassification.

## 1. INTRODUCTION

In a controlled, randomized experiment, treatment and control groups should be roughly the same—balanced—in their distributions of pretreatment variables. But how nearly so? Reports of clinical trials are urged to present tables of treatment and control group means of $x$-variables (Campbell et al., 2004), and they often do. These greatly assist qualitative assessments of similarity and difference between the groups, but in themselves they are silent as to whether, given the design, the discrepancies between the groups are large or small. Can the descriptive comparisons meaningfully be paired with significance tests? If so, must there be several, one for each variable, or can there be a single omnibus test? Would the omnibus test always require a simulation experiment, as proposed at some places in the literature on random assignment by group (Raab and Butcher, 2001)? Is there a large-sample test that is reliable in samples of moderate size, notwithstanding recent evidence to the contrary about one natural procedure (Gerber and Green, 2005)? At the level of foundations, some authors note that to assume experimental subjects to have been sampled from a superpopulation is antithetic to the nonparametric spirit common to randomized trials, and increasingly even to nonrandomized studies (Imai et al., 2008). Does testing for balance require a superpopulation-sampling model, as these authors also claim, or are there tests that more narrowly probe data's conformity to the experimental ideal? Relatedly, tests based on differences of group means require precise instructions for combining differences across strata or blocks, with the optimal approach appearing to depend on within- and between-stratum variation in $x$—within- and between-variation in the *population*, not the sample (Kalton, 1968). Does not the fine-tuning of these instructions require assumptions about, or estimation of, variability in the superpopulation, introducing sources of uncertainty that are generally ignored when drawing inferences about treatment effects (Yudkin and Moher, 2001)? Without notional superpopulations of $x$-values, how are alternatives to the null hypothesis to be conceived? What tests are optimal against these alternatives?

*Ben B. Hansen is Assistant Professor, Statistics Department, University of Michigan, 439 West Hall, Ann Arbor, Michigan 48109-1107, USA e-mail: ben.hansen@umich.edu. Jake Bowers is Assistant Professor, Political Science Department, University of Illinois at Urbana-Champaign, 361 Lincoln Hall, Urbana, Illinois 61801-3631, USA e-mail: jwbowers@uiuc.edu.*

The most familiar randomized comparisons of human subjects, perhaps, are drug and vaccine studies. Generally these are randomized at the level of individuals. But interventions upon neighborhoods, classrooms, clinics and families are increasingly the objects of study, and are increasingly studied experimentally; and even nonexperimental interventions at the group level may be analyzed using a combination of poststratification and analogies with hypothetical experiments. Might it be safe to ignore the group structure [as outcome analyses of cluster-randomized data often do (MacLennan et al. (2003); Isaakidis and Ioannidis (2003)), in some conflict with the recommendations of methodologists (Gail et al. (1996); Murray (1998); Donner and Klar (2000))] if interest focuses on individual-level outcomes, if correlations within group are low, or if the groups are small? Or, alternatively, do methods appropriate to individual-level assignment readily generalize to assignment by group?

### 1.1 Example: A Clinical Trial with Randomization at The Clinic Level

In order to study the benefit of up-to-date, best practices in monitoring and treatment of coronary heart disease, the ASSIST trial randomized 14 of 21 participating clinics to receive new systems for the regular review of heart disease patients (Yudkin and Moher, 2001). A primary outcome was whether monitoring assessments of heart patients met prescribed standards. One expects random assignment to make treatment and control clinics comparable in terms of what fractions of their heart patients were adequately assessed at baseline, and on baseline values of other relevant outcome variables. As is evident from Table 1, however, the clinics varied greatly in size and in patient characteristics; these differences limit the power of coin-tossing to smooth over preexisting differences. Seemingly sizable differences between treatment and control groups' proportions of adequately assessed patients may still compare favorably with differences that would have obtained in alternate random assignments. Viewed in isolation, such differences would appear, misleadingly, to threaten comparability of intervention groups. A principled means of distinguishing threatening and nonthreatening cases is needed.

A related need is for metrics with which to appraise the likely benefit, in terms of balance, of randomizing within blocks of relative uniformity on baseline measures.

### 1.2 Example: A Field Experiment on Political Participation

A second case in point is A. Gerber and D. Green's Vote'98 campaign, a voter turnout intervention in which get-out-the-vote (GOTV) appeals were randomly assigned to households of 1 or 2 voters. This is cluster-level randomization, because members of two-voter households were necessarily assigned to the same intervention; but with clusters containing no more than two individuals, it is as close to randomization of subjects as randomization of clusters can get. Accordingly, Gerber and Green's (2000) report gave outcome analyses that ignored clustering, effectively assuming their treatment assignments to have been independent of subjects', rather than clusters', covariates, and finding that in-person appeals effectively stimulated voting whereas solicitations delivered over the telephone, by professional calling firms, had little or no effect (Gerber and Green, 2000). Criticizing this analysis, Imai observes that the data Gerber and Green made available alongside their publication did not support the hypothesis of independence of subject-level covariates and treatment assignments (Imai, 2005). So poorly balanced are the groups, writes Imai, that the hypothesis of independence can be rejected at the $10^{-4}$ level (Imai (2005), Table 6). Had experimental protocol broken down, effectively spoiling the random assignment? Imai deduces that it must have, dismissing the original analysis and instead mounting another upon very different assumptions. Contrary to Gerber and Green, Imai's revision attaches significant benefits to paid GOTV calls.

In a pointed response, Gerber and Green (2005) shift doubt from the implementation of their experiment to Imai's methodology—particularly, the

TABLE 1
*Sizes of a subset of the 21 clinics participating in the ASSIST trial of register and recall systems for heart disease patients, along with baseline measurements of primary and secondary outcome variables*

| | Numbers of coronary heart disease patients | | | | |
| --- | --- | --- | --- | --- | --- |
| | | | Treated with | | |
| Practice # | In total | Adequately assessed | aspirin | hypotensives | lipid-reducers |
| 3  | 38  | 6  | 30  | 17 | 6  |
| 6  | 58  | 19 | 38  | 31 | 16 |
| 9  | 91  | 23 | 60  | 56 | 22 |
| 12 | 114 | 46 | 86  | 60 | 35 |
| 15 | 127 | 58 | 103 | 86 | 30 |
| 18 | 138 | 68 | 106 | 86 | 57 |
| 21 | 244 | 93 | 181 | 93 | 63 |

Despite the great variation in practice sizes, and in practice benchmarks targeted for improvement, a balanced allocation of practices to treatment conditions was sought. Adapted from Yudkin and Moher (2001), Table II.

method by which he checks for balance. Their counterattack has three fronts. First, they point out that Imai's analysis assumed independent assignment of individuals, whereas assignment really occurred at the household level. Second, they present results from a replication of the telephone GOTV experiment on a much larger scale, now randomizing individuals rather than households. The replication results were consistent with those of the original study. Third, they present simulation evidence that would cast doubt on Imai's recommended balance tests even had randomization been as he assumed. Those tests carried an asymptotic justification, for which the Vote'98 sample appears to have been too small—even though it comprised some 31,000 subjects, in more than 23,000 households!

The manifold nature of this argument makes methodological lessons difficult to draw. If the conclusion that the Vote'98 treatment assignment lacked balance is mistaken, then did the mistake lie in the conflation of household- and individual-level randomization, in the use of an inappropriate statistical test, or both?

### 1.3 Structure of the Paper

This and Section 2 introduce the paper. Section 3 develops randomization's consequences for the adjusted and unadjusted differences of group on baseline variables. Section 4 adapts these measures to testing for balance on several variables simultaneously. Section 5 develops theoretical arguments for the optimality of a specific approach recommended in Sections 3 and 4, and for the setting of a tuning constant, while Section 6 illustrates uses of the methodology for design and analysis. Section 7 concludes.

## 2. TWO WAYS NOT TO CHECK FOR BALANCE

This section examines appealing but ad hoc adaptations of two standard techniques, the method of standardized differences and goodness-of-fit testing with logistic regression, to the problem of testing for balance after random assignment of groups. To illustrate, we use the rich and publicly available Vote'98 dataset (Gerber and Green, 2005). It describes some 31,000 voters, falling in about 23,000 households; to complement this unusually large randomized experiment with a smaller one, we consider a simple random subsample of 100 households, comprising 133 voters. We study the association of the treatment assignment, $z$, with available covariates, $\mathbf{x}$, including age, ward of residence, registration status at the time of the previous election, whether a subject had voted in that election, and whether he had declared himself a member of a major political party. Telephone reminders to vote were attempted to roughly a fifth of the subjects, and it is around the putative randomness of this treatment assignment that Gerber, Green and Imai's debate centers.

### 2.1 Blurring the Difference Between Units of Assignment and Units of Measurement

Let us contrast *measurement units*, *subjects* or *elements*, here voters, with *clusters* or *assignment units*, here households containing one or two voters.

The *standardized difference* of measurement units on $x$ is a scaled difference of the average of $x$-values among measurement units in the treatment group and the corresponding average for controls. To facilitate interpretation, the difference is scaled by the reciprocal of one s.d. of measurement $x$'s, so that $100\times$ (standardized difference) can be read as a percent fraction of an s.d.'s difference. The purpose of this scaling, which is common in the matching literature (Cochran and Rubin (1973), page 420), is to standardize across $x$-variables; it differs from *direct standardization* of means or rates of disparate populations (cf., e.g., Fleiss (1973); Breslow and Day (1987)), in which subpopulations' means or rates are combined using a standard set of reference weights.

Setting the scaling aside, one has differences $\bar{x}_t - \bar{x}_c$, or, in vector notation, $\mathbf{z}^t\mathbf{x}/\mathbf{z}^t\mathbf{1} - (1-\mathbf{z})^t\mathbf{x}/(1-\mathbf{z})^t\mathbf{1}$, where $\mathbf{z} \in \{0,1\}^n$ indicates assignment to the treatment group. Considering this difference as a random variable, $\mathbf{Z}^t\mathbf{x}/\mathbf{Z}^t\mathbf{1} - (1-\mathbf{Z})^t\mathbf{x}/(1-\mathbf{Z})^t\mathbf{1}$, and conditioning on the numbers of measurement units in the treatment and control groups, $m_t = \mathbf{Z}^t\mathbf{1}$ and $m_c = (\mathbf{1}-\mathbf{Z})^t\mathbf{1}$, makes it a shifted random sum:

$$(1) \quad \frac{\mathbf{Z}^t\mathbf{x}}{m_t} - \frac{(\mathbf{1}-\mathbf{Z})^t\mathbf{x}}{m_c} = \mathbf{Z}^t\mathbf{x}/h - \mathbf{1}^t\mathbf{x}/m_c,$$

where $h = (m_c^{-1} + m_t^{-1})^{-1}$ is half of the harmonic mean of $m_c$ and $m_t$. Were treatment-group measurement units a simple random subsample of the sample as a whole, basic theory of simple random sampling would imply that (1) has mean zero and variance equal to $(m_t m_c/m) s^2(\mathbf{x})$, for $s^2(\mathbf{x}) = (m-1)^{-1}\sum_i (x_i - \bar{x})^2$ and $m = m_t + m_c$.

Consider instead the case in which a treatment group is selected by drawing a simple random sample of *clusters* of measurement units, but the analysis adopts the simplifying pretense that the group assigned to treatment constitutes a simple random sample of measurement units themselves. With this "fudge," differences $\bar{x}_t - \bar{x}_c$ are readily converted to $z$-scores. In the debate described in Section 1.2 above, both Gerber and Green (2000) and Imai (2005) took such an approach, perhaps reasoning that with cluster sizes no larger than two, differences between cluster- and individual-level randomization should be inconsequential.

We mounted a simulation experiment to determine whether this is so. The simulation mimicked the structure of the experiment's actual design, forming simulated treatment groups from random samples of 5275 of the 23,450 households, calculating differences $d_x^*$ in means of measurement unit $x$-values in the simulated treatment and control groups, and comparing these differences to the analogous difference $d_x$ between subjects to whom the Vote'98 campaign did and did not attempt a GOTV call. It reshuffled the treatment group $10^6$ times, making simulation $p$-values accurate to within 0.001. These $p$-values are given in the third and sixth columns of Table 2, which also presents $p$-values corresponding to the $z$-scores discussed above—$p$-values which ignore clustering—as well as large-sample $p$-values that account for clustering (by the method of Section 3.1, which attends to the difference of means of clusters' aggregated $x$-values rather than the difference of individuals' mean $x$-values). All $p$-values in Table 2 are two-sided.

The approximation ignoring the clustered nature of the randomization is not particularly good, especially for $m = 133$. Its $p$-values differ erratically from the actual $p$-values, at some points incorrectly suggesting departures from balance and elsewhere exaggerating it. (We had expressed the nominal "Ward" variable as 29 indicator variables, one for each ward, and the age measurement in terms of cubic B-splines with knots at quintiles of the age distribution, to yield six new $x$-variables; Table 2 displays the four of the 29 ward indicators, and the four of the six spline basis variables, for which the approximate $p$-values ignoring groups were most and least discrepant from actual $p$-values in the subsample and the full sample.) Increasing the sample size from 133 to 31,000 appears to improve the approximation somewhat, but not nearly as much as does explicitly accounting for clustering. It is noteworthy that pretending assignment was at the individual level leads to such striking errors—even with only half the experimental subjects assigned as part of a cluster, and even with no clusters larger than two.

### 2.2 The $p$-Value from Logistic Regression of Treatment Assignment on $x$'s

With or without treatment assignment by clusters, and with or without analytic adjustments to account for clusters, the method of standardized differences has the limitation that it produces a long list of nonindependent $p$-values, one for each $x$-variate studied. In many settings just a few $p$-values, ideally one, would be more convenient. This is true both when appraising the integrity of a randomization procedure, as in Imai (2005) sought to do, and poststratifying an observational study with goal of creating poststrata that resemble blocks of a randomized study in terms of observed covariates: for appraising such a poststratification, a list of possibly correlated test statistics is less helpful than a single omnibus test.

Logistic regression seems well suited to these tasks, particularly when treatment has been assigned at the measurement-unit level. For simple randomization, regress treatment assignment, $z$, on covariates $x$ and a constant, then on the constant alone, and compare the two fits using a standard asymptotic likelihood-ratio test. This one test speaks to whether $x$-variables influence $Z$, allowing each of the covariates to contribute to its verdict. Should the asymptotics of this deviance test apply, it will reject (at the 0.05 level) no more than about 5% of treatment assignments, presumably the ones in which, by coincidence, covariate balance failed to obtain. (For block randomization, the analogous approach involves regressing $z$ on $x$'s and a separate constant for each block, then on those constants alone.) There are problems with this procedure, however. Sample-size requirements are more stringent than one might think, are difficult to ascertain, and are typically incompatible with checking for balance thoroughly.

Table 3 shows the small-sample performance of the logistic regression deviance test, presenting the actual sizes of asymptotic-level 0.001, 0.01, 0.05 and 0.10 tests as applied to assignments of 14 of Yudkin and Maher's 21 clinics to treatment. The test's Type I error rates are markedly too high. Perhaps poor performance of asymptotic tests is to be expected, given the small sample size; but it is noteworthy that another asymptotic test, Section 4's method of combined baseline differences, succeeds in maintaining sizes no greater than advertised levels of significance.

Figure 1 illustrates the limited accuracy of the logistic regression approach in samples of moderate size. It compares asymptotic and actual null distributions of $p$-values from the logistic regression deviance test, effecting the actual distribution by simulation. One thousand simulation replicates are shown, both for the 100-household Vote'98 subsample and for the full sample. The covariates $x_{(1)}, \ldots, x_{(k)}$ are those described in Section 2.1, with $x$-values for two-person households determined by summing $x$-values of individuals in each household.

While $p$-values based on the asymptotic approximation appear accurate for the full sample, with

TABLE 3
*Small-sample ($n = 21$) Type I error rates of two types of test, one based on logistic regression and another, to be described in Section 4, based on adjusted differences of treatment and control groups' covariate means*

| | Size of test | | | |
|---|---|---|---|---|
| | Asymptotic | | | |
| Method | 0.001 | 0.01 | 0.05 | 0.10 |
| | Actual | | | |
| Logistic regression-based | 0.0281 | 0.0620 | 0.16 | 0.24 |
| Combined baseline differences | 0.0000 | 0.0003 | 0.018 | 0.064 |

The actual size of the logistic regression tests well exceeds their nominal levels, while the alternate test is somewhat conservative but holds to advertised levels. Based on $10^6$ simulated assignments to treatment of 14 of the 21 ASSIST clinics.

TABLE 2
*Effect of accounting for assignment by groups on approximations to p-values, in the full Vote'98 sample and in a subsample of 100 households*

| | 100 households ($m = 133$) | | | All households ($m = 31$K) | | |
|---|---|---|---|---|---|---|
| | Accounting for groups? | | | Accounting for groups? | | |
| Baseline variable ($x$) | No | Yes | Actual | No | Yes | Actual |
|---|---|---|---|---|---|---|
| Number of voters in household | 0.12 | 0.24 | 0.21 | 0.85 | 0.82 | 0.82 |
| Voted in 1996 | 0.40 | 0.22 | 0.22 | 0.23 | 0.39 | 0.39 |
| Major party member | 0.45 | 0.14 | 0.16 | 0.24 | 0.18 | 0.18 |
| Bspline$_2$(Age) | 0.68 | 0.59 | 0.60 | 0.06 | 0.31 | 0.31 |
| Bspline$_4$(Age) | 0.82 | 0.39 | 0.40 | 0.68 | 0.68 | 0.68 |
| Bspline$_5$(Age) | 0.72 | 0.24 | 0.24 | 0.39 | 0.22 | 0.22 |
| Bspline$_6$(Age) | 0.19 | 0.62 | 0.62 | 0.56 | 0.89 | 0.89 |
| Ward 2 | 1.00 | 1.00 | 0.50 | 0.81 | 0.87 | 0.87 |
| Ward 5 | 0.89 | 0.98 | 0.89 | 0.44 | 0.47 | 0.48 |
| Ward 10 | 0.58 | 0.54 | 0.65 | 0.95 | 0.97 | 0.97 |
| Ward 11 | 0.75 | 0.92 | 0.87 | 0.27 | 0.42 | 0.42 |

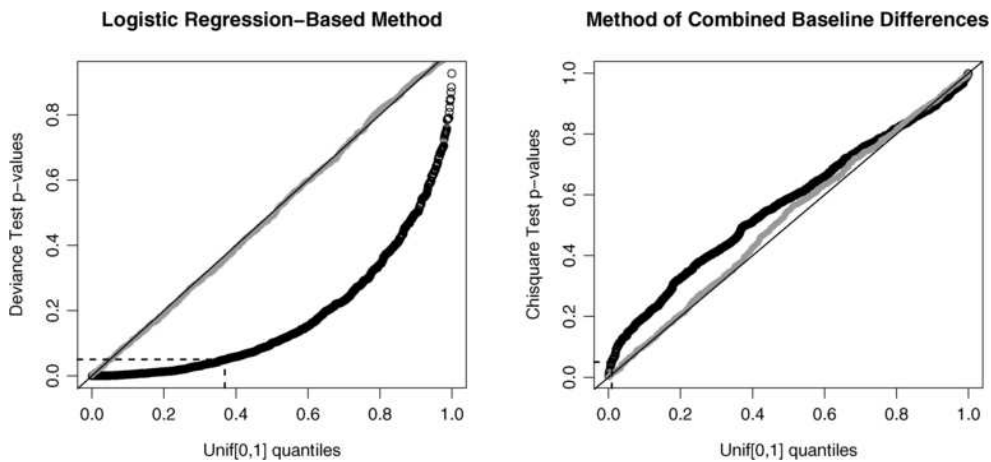

FIG. 1. *Theoretical and actual p-values of two omnibus tests of covariate balance, both accounting for clustering. With 100 assignment units and 38 degrees of freedom (dark trace), logistic regression's p-values are markedly too small, whereas p-values from the method of combined baseline differences (Section 4) err toward conservatism, and to a lesser degree. The dashed lines at left indicate that the logistic regression-based test yielded p-values less than $0.05$ in just under 40% of simulated random assignments, whereas the dashed lines at right indicate that the combined baseline difference-test with nominal level $\alpha = 0.05$ had actual size of about $0.01$. However, with the full 23,000 assignment units (and the same 38 degrees of freedom), both methods perform as their asymptotics would predict, as indicated by the close agreement in both panels of the lighter traces and the 45° lines.*

its 23 thousand-someodd households, those for the subsample are quite exaggerated. In it, the nominal 0.05-level test has an actual size of about 0.37. Would an alert applied statistician have identified the 100-household subsample as too small for the likelihood ratio test? Perhaps; it has only $2\frac{1}{2}$ times as many observations as $x$-variables, once the Age and Ward variables have been expanded as in Table 2. But how large a ratio of observations to covariates would be sufficient? Intuition may be a poor guide. To explore the difference in information carried by binary and continuous outcomes, Brazzale, Davison and Reid (2006, Section 4.2; see also Davison (2003), ex. 10.17) construct artificial data sets from a real one with a binary independent variable, some retaining the binary outcome structure but increasing the apparent information in the dataset by replicating observations, and others imputing continuous outcomes according to a logistic distribution. Their results are striking; one observation with continuous response carries about as much information as eight observations with binary response, and deviance tests are found to be unreliable even with 11 times as many observations as $x$-variables. Harrell (2001, Section 4.5), Peduzzi et al. (1996) and Whitehead (1993) offer somewhat less pessimistic guidelines, but even these would require well more than 10 times as many observations as $x$-variables— an odd condition to place on a comparative study, one which many otherwise strong studies would violate.

For contrast, the right panel of Figure 1 offers an analogous comparison between asymptotically approximate and actual $p$-values of a test statistic to be introduced in Section 4. Even with relatively few observations as compared to $x$-variables, its size never exceeds its nominal level (if it errs somewhat toward conservatism).

## 3. RANDOMIZATION TESTS OF BALANCE, WITH AND WITHOUT CLUSTERS

A common form of frequentism, sometimes traced to Neyman (1923), posits that subjects arrive in a study through random sampling from a broader population, and takes as its goal to articulate characteristics of that population. An impediment to applying this conceptualization to comparative studies is that their samples need not represent background populations. Comparing within a sample and extrapolating from it are separate goals, neither of which needs to depend on the other. In contrast, in Fisher's model of a comparative study no background population is supposed, but randomization is supposed to govern division of the sample into comparison groups. Inference asks what differences between groups can be explained by chance, rather than what differences between sample and population can be explained by chance. Fisher's approach is better suited to appraising balance.

### 3.1 Experiments with Simple Randomization of Clusters

To illustrate, consider the question of whether in the Vote'98 experiment subjects assigned to receive a telephone reminder had voted in the prior election in similar proportion to those not so assigned. Since past voting is predictive of future voting, sizable differences to the advantage of either group may cause estimates of treatment effects to err, reflecting the baseline difference more than effects of GOTV interventions.

Let the index $i = 1, \ldots, n$ run over assignment units, so that $z_i$ indicates the treatment assignment of the $i$th cluster of observation units. Interpret $x_i$ as the total of $x$-values for observation units in cluster $i$, in this case the number of subjects in the household who voted in the previous election, and let $m_i$ be the size of that cluster, here 1 or 2, in observation units. $\mathbf{z}$, $\mathbf{x}$ and $\mathbf{m}$ are $n$-vectors recording these data for each assignment unit. The observed difference of the proportions of treatment and control group subjects who had cast votes in 1996 can be written as a function $d_p(\mathbf{z}, \mathbf{x})$ of the treatment-group indicator vector $\mathbf{z}$ and indicators $\mathbf{x}$ of voting in the previous election. In symbols, $d_p(\mathbf{z}, \mathbf{x}) = \mathbf{z}^t\mathbf{x}/\mathbf{z}^t\mathbf{m} - (\mathbf{1} - \mathbf{z})^t\mathbf{x}/(\mathbf{1} - \mathbf{z})^t\mathbf{m}$; for general measurement variables $v$, $d_p(\mathbf{z}, \mathbf{v})$ is the difference of treatment and control group means. Let $A$ be the set of treatment assignments from which the actual assignment $\mathbf{z}$ was randomly selected; for each member $\mathbf{z}^*$ of $A$, it is straightforward to compute the amount $d_p(\mathbf{z}^*, \mathbf{x})$ by which treatments and controls would have differed had assignment $\mathbf{z}^*$ been selected. A (two-sided) randomization $p$-value attaching to the hypothesis of nonselection on $\mathbf{x}$ is

$$
\begin{aligned}
&\frac{\#\{\mathbf{z}^* \in A : |d_p(\mathbf{z}^*, \mathbf{x})| > |d_p(\mathbf{z}, \mathbf{x})|\}}{\#A} \\
&\quad + \frac{(1/2)\#\{\mathbf{z}^* \in A : |d_p(\mathbf{z}^*, \mathbf{x})| = |d_p(\mathbf{z}, \mathbf{x})|\}}{\#A} \\
&= \mathbf{P}(|d_p(\mathbf{Z}, \mathbf{x})| > |d_p(\mathbf{z}, \mathbf{x})|) \\
&\quad + \frac{1}{2}\mathbf{P}(|d_p(\mathbf{Z}, \mathbf{x})| = |d_p(\mathbf{z}, \mathbf{x})|),
\end{aligned}
\tag{2}
$$

where $\mathbf{Z}$ is a random vector that is uniformly distributed on possible treatment assignments $A$. (Weighting by one-half those $\mathbf{z}^* \in A$ for which $|d_p(\mathbf{z}^*, \mathbf{x})| = |d_p(\mathbf{z}, \mathbf{x})|$ makes this a mid-$p$ value, the null distribution of which will be more nearly uniform on $[0,1]$ than would a $p$-value without this weighting. Agresti and Gottard (2005) discuss merits of the mid-$p$ value.) This appraisal of balance on $x$ does involve probability, but only treatment assignment, not the covariate, is modeled as stochastic.

In principle, these $p$-values can be determined exactly, perhaps by enumeration; in practice, it is accurate enough, and often much easier, to evaluate them by simulation [as does, e.g., Lee (2006)]. Under favorable designs, fast and accurate Normal approximations are also available. Consider first the case in which

(A) the assignment scheme allocates a fixed and predetermined number $n_t$ of the $n$ clusters to treatment, and
(B) each cluster contains the same number $m_0$ of measurement units.

Then the ratios $\mathbf{Z}^t\mathbf{x}/\mathbf{Z}^t\mathbf{m}$ and $(\mathbf{1} - \mathbf{Z})^t\mathbf{x}/(\mathbf{1} - \mathbf{Z})^t\mathbf{m}$ of which $d_p(\mathbf{Z}, \mathbf{x})$ is a difference have constants, respectively, $k_0 = m_0 n_t$ and $k_1 = m_0 n_c$, as denominators, so that, as in (1), $d_p(\mathbf{Z}, \mathbf{x})$ has an equivalent of the form $\mathbf{Z}^t\mathbf{x}/k_0 - \mathbf{1}^t\mathbf{x}/k_1$. Then it is necessary only to approximate the distribution of $\mathbf{Z}^t\mathbf{x}$, an easier task than approximating the distribution of its ratio with another random variable. Indeed, if $\{i \in \{1, \ldots, n\} : Z_i = 1\}$ is a simple random sample of size $n_t$, then $\mathbf{Z}^t\mathbf{x}$ is simply the sample sum of a simple random sample of $n_t$ from $n$ cluster totals $x_1, \ldots, x_n$. Common results for simple random sampling give that $\mathbf{E}(\mathbf{Z}^t\mathbf{x}) = n_t \bar{x} = \frac{n_t}{n}\sum x_i$; that $\mathrm{Var}(\mathbf{Z}^t\mathbf{x}) = n_t(1 - \frac{n_t}{n})s^2(\mathbf{x})$, where $s^2(\mathbf{x}) = (\sum_1^n (x_i - \bar{x})^2)/(n-1)$; and that if $x$ has few or no outliers and is not particularly skewed, then if $n$ is sufficiently large and $n_t/n$ is neither near 0 nor 1, the law of $\mathbf{Z}^t\mathbf{x}$ will be roughly Normal. [Formally, if $n_t$ grows to infinity while $n_t/n$ approaches a constant in $(0, 1)$, and mean squares and cubes of $|\mathbf{x}|$ remain bounded, then the limiting distribution of $\mathbf{Z}^t\mathbf{x}$ is Normal (Hájek (1960); Erdős and Rényi (1959)).] Over and above this finite population central limit theorem, Höglund's Berry–Esseen principle for simple random sampling (Höglund (1978)) limits the error of the Normal approximation in finite samples, suggesting that it should govern $\mathbf{Z}^t\mathbf{v}$ similarly well for well behaved covariates $v$ other than $x$, and that it should be quite good even in samples of moderate size. Note that covariates $x$ which are ill-behaved, in the sense of being skewed or having extreme outliers, are also ill-suited to be summarized in terms of their means in any event—thus transformations to more regular covariates $\tilde{x} = f(x)$ are advisable in order to ease description, regardless of differences between $d(\mathbf{Z}^t, \mathbf{x})$'s and $d(\mathbf{Z}^t, \tilde{\mathbf{x}})$'s sampling properties; and insofar as $\bar{\tilde{x}}$ appropriately measures the central tendency of $(\tilde{x}_i : i \leq n)$, $\tilde{x}$ will be well-behaved in the sense needed for $d(\mathbf{Z}^t, \tilde{\mathbf{x}}) \sim \mathcal{N}(n_t \bar{\tilde{x}}, n_t(1 - \frac{n_t}{n}) s^2(\tilde{\mathbf{x}}))$.

Cases in which (A) or (B) fails might appear to frustrate this argument. For instance, suppose treatment were assigned, in violation of (A), by $n$ independent Bernoulli$(p)$ trials. Then there would be some random fluctuation in treatment and control group sizes $\mathbf{Z}^t \mathbf{m}$ and $(\mathbf{1} - \mathbf{Z})^t \mathbf{m}$, and the denominators of the ratios of which $d_p(\mathbf{Z}, \mathbf{x})$ is a difference would no longer be constants, so that the argument by which Höglund's Berry–Esseen principle bounded the error of the Normal approximation would no longer be available. However, this particular frustration is circumvented by referring observed differences $d_p(\mathbf{z}, \mathbf{x})$ to conditional, rather than marginal, distributions of $d_p(\mathbf{Z}, \mathbf{x})$. For conditional on $\mathbf{Z}^t \mathbf{1} = \mathbf{z}^t \mathbf{1} = n_t$, condition (A) is restored, and provided (B) also holds the distribution of $d_p(\mathbf{Z}, \mathbf{x})$ is close to Normal, with mean and variance as previously indicated.

What of departures from (B), that is, clusters that vary in size? Here the representation of $d_p(\mathbf{Z}, \mathbf{x})$ as a linear transformation of $\mathbf{Z}^t \mathbf{x}$ need not apply, even after conditioning on the number of clusters selected for treatment, since then the number of treatment-group subjects $\mathbf{Z}^t \mathbf{m}$ may vary between possible assignments. A modification to $d_p(\cdot, \cdot)$ circumvents the problem. Now writing $m_t$ for the expected, rather than observed, number of measurement units in the treatment group, set

$$d(\mathbf{z}, \mathbf{x}) := \frac{\mathbf{z}^t \mathbf{x}}{m_t} - \frac{(\mathbf{1} - \mathbf{z})^t \mathbf{x}}{m - m_t}$$
$$[m_t := \mathbf{E}(\mathbf{Z}^t \mathbf{m}), m = \mathbf{1}^t \mathbf{m}]$$
$$= \bar{m}^{-1} [\mathbf{z}^t \mathbf{x}/h - \mathbf{1}^t \mathbf{x}/(n - n_t)]$$
$$[h := [n_t(1 - n_t/n)]^{-1}].$$

Kerry and Bland (1998) recommend an analogous statistic for outcome analysis in cluster randomized trials.

In designs with size variation among assignment units, $d(\mathbf{z}, \mathbf{x})$ and $d_p(\mathbf{z}, \mathbf{x}) = \mathbf{z}^t \mathbf{x}/\mathbf{z}^t \mathbf{m} - (\mathbf{1} - \mathbf{z})^t \mathbf{x}/(m - \mathbf{z}^t \mathbf{m})$ may differ. The differences will tend to be small, particularly if $\mathbf{m}$, now regarded as a covariate, is well balanced; and of course this balance is expediently measured using $d(\mathbf{z}, \mathbf{m})$ and its associated $p$-value.

These considerations recommend $d(\mathbf{z}, \mathbf{x})$ as a basic measure of balance on a covariate $x$.

### 3.2 Simple Randomization of Clusters within Blocks, Strata or Matched Sets

The approach extends to the case of block-randomized designs, and to designs that result from post-stratification or matching. Let there be strata $b = 1, \ldots, B$, within which simple random samples of $n_{t1}, \ldots, n_{tB}$ clusters are selected into the treatment group from $n_1, \ldots, n_B$ clusters overall, for each $b = 1, \ldots, B$. Let $\mathbf{Z} = (\mathbf{Z}_1^t, \ldots, \mathbf{Z}_b^t, \ldots, \mathbf{Z}_B^t)^t$, $\mathbf{Z}_b^t = (Z_{b1}, \ldots, Z_{bn_b})$ for each stratum $b$, be a vector random variable of which the experimental assignment was a realization, and let $\mathbf{m} = (\mathbf{m}_1^t, \ldots, \mathbf{m}_b^t)^t$ record sizes of clusters in terms of observation units. For each $b = 1, \ldots, B$, let $m_{tb} = \mathbf{E}(\mathbf{Z}_b^t \mathbf{m}_b) = \bar{m}_b n_{tb}$ be the expected number of observation units in the treatment group. Let $\mathbf{x} = (\mathbf{x}_1^t, \ldots, \mathbf{x}_B^t)^t$ and $\mathbf{v} = (\mathbf{v}_1^t, \ldots, \mathbf{v}_B^t)^t$ be single covariates—perhaps cluster sums of individual measurements.

Because both treatment "propensities" [i.e., $\mathbf{P}(Z_{b1} = 1)$, $b = 1, \ldots, B$] and covariate distributions may vary across blocks, comparisons of simple means of treatment and control units, even assignment units rather than measurement units, may fall prey to Simpson's paradox, despite random assignment (Blyth, 1972). Rather, when averaging across blocks the two means must be standardized by a common set of block-specific weights; or, equivalently, treatment and control averages can be taken and compared within blocks before taking the weighted average of the differences. Within a block $b$, the (modified) difference of treatment and control group means on $\mathbf{x}$ is simply $\mathbf{z}_b^t \mathbf{x}_b / m_{tb} - (\mathbf{1} - \mathbf{z}_b)^t \mathbf{x}_b / (m - m_{tb})$. Weights may be proportional to the number of subjects in each block, proportional to the number of treatment-group subjects in each block, or selected so as to be optimal under some model; this latter approach is developed in Section 5. For now, fix positive weights $w_1, \ldots, w_B$ such that $\sum_i w_i = 1$.

Considered as a random variable, the adjusted difference of treatment and control group means is

$$d(\mathbf{Z}, \mathbf{x}) = \sum_{b=1}^{B} w_b [\mathbf{Z}_b^t \mathbf{x}_b / m_{tb}$$

(3)
$$- (\mathbf{1} - \mathbf{Z}_b)^t \mathbf{x}_b / (m - m_{tb})]$$

(4)
$$= \sum_{b=1}^{B} w_b h_b^{-1} \bar{m}_b^{-1} \mathbf{Z}_b^t \mathbf{x}_b$$
$$- \sum_{b=1}^{B} w_b \bar{m}_b^{-1} (n_b - n_{tb})^{-1} \mathbf{1}^t \mathbf{x}_b,$$

where $h_b = [n_{tb}^{-1} + (n_b - n_{tb})^{-1}]^{-1} = [n_{tb}(1 - n_{tb}/n_b)]$ is half the harmonic mean of $n_{tb}$ and $(n_b - n_{tb})$. Within block $b$, $\mathbf{Z}_b^t \mathbf{x}_b$ is the sample total of a simple random sample of size $n_{tb}$ from $(x_{b1}, \ldots, x_{bn})$. It follows that it has mean $(n_{tb}/n_b) \mathbf{1}^t \mathbf{x}_b = n_{tb} \bar{x}_b$; that its variance is $h_b s^2(\mathbf{x}_b)$; and that its covariance with $\mathbf{Z}_b^t \mathbf{v}_b$ is $h_b s(\mathbf{x}_b; \mathbf{v}_b)$, for $s(\mathbf{x}_b; \mathbf{v}_b) = (\mathbf{x}_b - \bar{x}_b \mathbf{1})^t (\mathbf{v}_b - \bar{v}_b \mathbf{1})/(n_b - 1)$ and $s^2(\mathbf{x}_b) = s(\mathbf{x}_b; \mathbf{x}_b)$. By virtue of the design, for blocks $b' \neq b$ the treatment-group totals $\mathbf{Z}_b^t \mathbf{x}_b$ and $\mathbf{Z}_b^t \mathbf{v}_b$ are independent of $\mathbf{Z}_{b'}^t \mathbf{x}_{b'}$ and $\mathbf{Z}_{b'}^t \mathbf{v}_{b'}$. Together, these facts entail the following description of the first and second moments of $d(\mathbf{Z}, \mathbf{x})$ and $d(\mathbf{Z}, \mathbf{v})$.

PROPOSITION 3.1. *Suppose that within blocks $b = 1, \ldots, B$, simple random samples of $n_{tb}$ from $n_b$ clusters are selected for treatment, with the rest assigned to control. Let $\mathbf{Z}$ indicate sample membership and let $\mathbf{x}$ and $\mathbf{v}$ denote covariates. For $d(\cdot, \cdot)$ as in (3), one has*

$$\mathbf{E}(d(\mathbf{Z}, \mathbf{x})) = \mathbf{E}(d(\mathbf{Z}, \mathbf{v})) = 0,$$
$$\text{Var}(d(\mathbf{Z}, \mathbf{x})) = \sum_{b=1}^{B} \frac{w_b^2}{h_b \bar{m}_b} \frac{s^2(\mathbf{x}_b)}{\bar{m}_b},$$
$$\text{Cov}(d(\mathbf{Z}, \mathbf{x}), d(\mathbf{Z}, \mathbf{v})) = \sum_{b=1}^{B} \frac{w_b^2}{h_b \bar{m}_b} \frac{s(\mathbf{x}_b; \mathbf{v}_b)}{\bar{m}_b},$$

*where $h_b = [n_{tb}^{-1} + (n_b - n_{tb})^{-1}]^{-1}$.*

When $d(\mathbf{Z}, \mathbf{x})$ can be assumed Normal, Proposition 3.1 permits analysis of its distribution. In fact, relevant central limit theorems do entail its convergence to the Normal distribution as the size of the sample increases, and they suggest that the convergence should be fast and uniform across covariates $\mathbf{x}, \mathbf{v}, \ldots$. There are two cases. In the first case, the size of each stratum falls under a fixed limit. Since the sample size is increasing, this means the number of strata tends to infinity. As each of them makes an independent contribution to the sum that is $d(\mathbf{Z}, \mathbf{x})$, ordinary central limit theorems entail that its distribution tends to Normal. Indeed, the ordinary Berry–Esseen lemma limits the difference between the distribution function of $d(\mathbf{Z}, \mathbf{x})$ and an appropriate Normal distribution in terms of its variance and its third central moment (Feller 1971, Chapter 16) both of which are calculable precisely from the design and from the configuration of $\mathbf{x}$. In the second case, at least one stratum size tends to infinity. Assume that in each growing stratum the proportions of clusters assigned to treatment and to control tend to nonzero constants. Then the contribution $(h_b \bar{m}_b)^{-1} \mathbf{Z}_b^t \mathbf{x}_b$ from any growing stratum $b$ is a rescaled sum of a simple random sample from $(x_{b1}, x_{b2}, \ldots, x_{bn_b})$ and is governed by the central limit theorem and Berry–Esseen principle for simple random sampling (see Section 3.1). Contributions from small strata that do not grow are either asymptotically Normal, by the first argument, or, assuming a nonpathological weighting scheme, asymptotically negligible, or both; it follows that the overall sum of stratum contributions tends to Normal.

Although any weighting of blocks is possible, some are more likely to reveal imbalances than others. Section 5 shows weighting in proportion to the product of block-mean cluster size and the harmonic mean of $n_{tb}$ and $n_b - n_{tb}$, $w_b^* \propto h_b \bar{m}_b$, to be optimal in an important sense. It also so happens that with this weighting, expressions for the first and second moments of $d(\mathbf{Z}, \mathbf{x})$ simplify.

COROLLARY 3.1. *Suppose that within blocks $b = 1, \ldots, B$, simple random samples of $n_{tb}$ from $n_b$ clusters are selected for treatment, with the rest assigned to control. Let $\mathbf{Z}$ indicate sample membership and let $\mathbf{x}$ and $\mathbf{v}$ denote covariates. For $d(\cdot, \cdot)$ as in (3), with $w_b \equiv w_b^* \propto h_b \bar{m}_b = \bar{m}_b n_{tb}(1 - n_{tb}/n_b)$, one has*

(5)
$$d(\mathbf{z}, \mathbf{x}) = \left( \sum h_b \bar{m}_b \right)^{-1}$$
$$\cdot \left[ \sum_{b=1}^{B} \mathbf{Z}_b^t \mathbf{x}_b - \sum_{b=1}^{B} n_{tb} (\mathbf{1}^t \mathbf{x}_b / n_b) \right],$$
$$\mathbf{E}(d(\mathbf{Z}, \mathbf{x})) = \mathbf{E}(d(\mathbf{Z}, \mathbf{v})) = 0,$$

(6)
$$\text{Var}(d(\mathbf{Z}, \mathbf{x})) = \left( \sum h_b \bar{m}_b \right)^{-2}$$
$$\cdot \sum_{b=1}^{B} h_b \bar{m}_b \frac{s^2(\mathbf{x}_b)}{\bar{m}_b},$$

$$\mathrm{Cov}(d(\mathbf{Z},\mathbf{x}), d(\mathbf{Z},\mathbf{v})) = \left(\sum h_b \bar{m}_b\right)^{-2}$$
$$\cdot \sum_{b=1}^{B} h_b \bar{m}_b \frac{s(\mathbf{x}_b; \mathbf{v}_b)}{\bar{m}_b}.$$

### 3.3 Accommodating Independent Assignment by Conditioning

Proposition 3.1 assumes simple random sampling of treatment groups within blocks. Were assignments within block $b$ made by independent Bernoulli($p_b$) trials, the induced first and second moments of $d(\mathbf{Z},\mathbf{x})$—understood as a $w_b$-weighted sum of terms

$$\frac{\mathbf{Z}_b^t \mathbf{x}_b}{\bar{m}_b \mathbf{Z}_b^t \mathbf{1}} - \frac{(\mathbf{1} - \mathbf{Z}_b)^t \mathbf{x}_b}{\bar{m}_b (n_b - \mathbf{Z}_b^t \mathbf{1})},$$

since $n_{tb}$ would no longer be a fixture of the design—would be formally and numerically similar to those of the proposition, as a simple argument shows. $\mathbf{Z}_b^t \mathbf{1}$ is $\mathrm{Bin}(n_b, p_b)$, independently of $\mathbf{Z}_{b'}^t \mathbf{1} \sim \mathrm{Bin}(n_{b'}, p_{b'})$, $b' \neq b$, and conditionally on $\mathbf{Z}_b^t \mathbf{1} = n_{bt}$ the distribution of $\mathbf{Z}_b^t \mathbf{x}_b$ is that of a sample sum of a simple random sample of size $n_t$ from $\{x_{b1}, \ldots, x_{bn_b}\}$. In general, conditioning on $\mathbf{Z}_1^t \mathbf{1}, \ldots, \mathbf{Z}_B^t \mathbf{1}$ gives $d(\mathbf{Z},\mathbf{x})$ and $d(\mathbf{Z},\mathbf{v})$ distributions of the type described in Proposition 3.1.

Conditional assessments of $d(\mathbf{Z},\mathbf{x})$ have the advantage of being immune from disruption by unusually small or large allocations $\mathbf{Z}_i^t \mathbf{1}$ to treatment, $i = 1, \ldots, B$. The sizes of these allocations carry little relevant information, as a conditionality argument shows. Consider the broader model in which $\mathbf{P}(Z_{bi} = 1)$ is not a constant for all $i = 1, \ldots, n_b$, but instead $\mathrm{logit}(\mathbf{P}(Z_{bi})) = \psi_b + \psi_x(x_{bi})$. The null hypothesis holds that $\psi_x \equiv 0$; a test of balance aims to reject it when $\psi_x(\cdot)$ is nonnull. The likelihood of the full model, with independent sampling of $Z_{bi}$'s and possibly nonzero $\psi_x$, can be straightforwardly represented as

$$(7) \quad \prod_b \exp\left\{ \left(\sum_{i=1}^{n_b} Z_{bi}\right) \psi_b + \sum_{i=1}^{n_b} Z_{bi} \psi_x(x_{bi}) - k_b(\psi_b, \psi_x) \right\},$$

$k_b = \sum_{i=1}^{n_b} \log[1 + \exp(\psi_b + \psi_x(x_{bi}))]$, but it can also be parametrized in terms of the function $\psi_x(\cdot)$ and moment parameters $\eta_b = \mathbf{E}(\mathbf{Z}_b^t \mathbf{1} | \psi_b, \psi_x)$, $b = 1, \ldots, B$, with $(\eta_1, \ldots, \eta_B)$ and $\psi_x(\cdot)$ being variation independent (Barndorff-Nielsen and Cox (1994), page 40 ff). The statistic $(\mathbf{Z}_1^t \mathbf{1}, \ldots, \mathbf{Z}_B^t \mathbf{1})$ is ancillary for inference about the function $\psi_x$; in the main it reflects on $(\eta_b : b \leq B)$, not $\psi_x(\cdot)$.

### 3.4 Example and Implementation

The Vote'98 experiment used a factorial design, varying the probability of a household's assignment to telephone GOTV across levels of the other treatments it assigned. (Specifically, households eligible for telephone calls were also eligible for assignment to receive GOTV mailings, and for assignment to receive a personal GOTV appeal; the probability of assignment to the telephone group varied across cells of the mail GOTV by personal GOTV cross-classification.) This makes methods for block-randomized studies a necessity. Consequently, Table 2 uses modified differences of Section 3.2, as aggregated using harmonic block weights as in Corollary 3.1, to combine balance measures across subclasses defined by treatments other than telephone GOTV.

The first row of Table 2 gives results for the test as to whether $\mathbf{z}^t \mathbf{m}$, the size of the treatment group in measurement units, differed substantially from $\mathbf{E}(\mathbf{Z}^t \mathbf{m})$, in a subsample of 100 clusters and in the full sample of some 23,000. The $z$-scores $d(\mathbf{z}, \mathbf{m}) / \sqrt{V(d)}$ (not shown in the table) were 1.186 and 0.226 for the sub- and full samples, respectively, which by Normal tables give approximate $p$-values of 0.236 and 0.821. This suggests $\mathbf{z}^t \mathbf{m}$ was relatively quite close to its null expectation, a suggestion that gains further support from simulations, which find the mid-$p$ values to be 0.211 and 0.821, respectively. Having confirmed balance on cluster sizes, the next row of the table asks about voting in the previous election. It is not precisely the same in treatment and control groups, either for the subsample or for the full sample, as indicated by normalized differences of $d(\mathbf{z}, \mathbf{x}) / \sqrt{V(d)} = 1.228$ and $-0.853$, respectively; but the $p$-values, 0.224 and 0.391, indicate that voting in the previous election is as similar in the two groups as could be expected from random assignment, and the Normal approximation locates them with some accuracy, 0.220 and 0.394.

To compute these adjusted baseline differences and their large-sample reference distributions, the first step, prior to calling any specialized function, is to aggregate the data to the cluster level, recording cluster sizes $m_{bi}$ and creating cluster totals $x_{bi}$ from individual measurements $x_{bi1}, \ldots, x_{bim_{bi}}$. R users can then adapt functionality from either of at least two R packages, Bowers and Hansen's (2006) RITOOLS or Hothorn et al.'s (2006) COIN, which perform randomization-based inference without explicit attention to cluster-level assignment. We give details for RITOOLS, which uses harmonic weights, $w_b \propto h_b$, by default. Its function `xBalance` calculates

$$
\begin{aligned}
(8) \quad & d_{\text{no clus}}(\mathbf{z}, \mathbf{x}) \\
&= \left(\sum_{b=1}^{B} h_b\right)^{-1} \\
&\quad \cdot \left\{\sum_{b=1}^{B} h_b[\mathbf{z}_b^t \mathbf{x}_b / n_{tb} \right. \\
&\quad \left. - (\mathbf{1} - \mathbf{z}_b)^t \mathbf{x}_b / (n_b - n_{tb})]\right\}
\end{aligned}
$$

and its randomization variance, printing significance stars based on the corresponding $z$-score, $\text{Var}(d_{\text{no clus}}(\mathbf{Z}, \mathbf{x}))^{-1/2} d_{\text{no clus}}(\mathbf{z}, \mathbf{x})$. [The $z$-score itself is not displayed; instead, as a descriptive measure `xBalance` reports a standardized difference in the sense of Section 2.1, namely $s_p^{-1}(\mathbf{x}) d_{\text{no clus}}(\mathbf{z}, \mathbf{x})$ where $s_p(\mathbf{x})$ is the pooled s.d. of $\mathbf{x}$ in the sense of the two-sample $t$-test comparing treatment to control clusters.] Since (8) differs from (3) with $w_b \propto h_b \bar{m}_b$ only in that its denominator is $\sum_{b=1}^{B} h_b$ rather than $\sum_{b=1}^{B} h_b \bar{m}_b$, this $z$-score is the same as that which Corollary 3.1 would have given.

With other software, $\text{Var}(d(\mathbf{Z}, \mathbf{x}))$ may have to be calculated explicitly from (6), but a shortcut is available for determining $d(\mathbf{z}, \mathbf{x})$ when $w_b \propto h_b \bar{m}_b$: $d_{\text{no clus}}(\mathbf{z}, \mathbf{x})$, or $[(\sum_b h_b \bar{m}_b)/(\sum_b h_b)] d(\mathbf{z}, \mathbf{x})$, coincides with the ordinary least squares coefficient of $\mathbf{z}$ in the regression of $\mathbf{x}$ on $\mathbf{z}$ and dummy variables for blocks. Unfortunately, $\text{Var}(d(\mathbf{Z}, \mathbf{x}))$ does not relate in any helpful way to this coefficient's ordinary least squares standard error. To recover $d(\mathbf{z}, \mathbf{x})$ from the least squares coefficient, $\sum_b h_b$ and $\sum_b h_b \bar{m}_b$ will have to be calculated. However, given that (6) has to be figured, these calculations pose little additional burden; and they are the same for each variable $\mathbf{x}$ on which balance is to be checked.

## 4. SIMULTANEOUSLY TESTING BALANCE ON MULTIPLE $x$'S

Ordinarily there will be several, perhaps many, $x$-variables along which balance ought to be checked, and a method of combining baseline differences will be needed. To this end, write

$$
\begin{aligned}
(9) \quad & d^2(\mathbf{z}; \mathbf{x}_1, \ldots, \mathbf{x}_k) \\
& := [d(\mathbf{z}, \mathbf{x}_1), \ldots, d(\mathbf{z}, \mathbf{x}_k)]
\end{aligned}
$$

$$
\cdot \left\{\text{Cov}\left(\begin{bmatrix} d(\mathbf{Z}, \mathbf{x}_1) \\ \vdots \\ d(\mathbf{Z}, \mathbf{x}_k) \end{bmatrix}\right)\right\}^{-} \begin{bmatrix} d(\mathbf{z}, \mathbf{x}_1) \\ \vdots \\ d(\mathbf{z}, \mathbf{x}_k) \end{bmatrix},
$$

where $\text{Cov}(d(\mathbf{Z}, \mathbf{x}_i), d(\mathbf{Z}, \mathbf{x}_j))$ is as in Proposition 3.1 and $M^-$ denotes a generalized inverse of $M$. This test has the desirable properties that: (i) it culminates in a single test statistic and $p$-value; (ii) its law is roughly $\chi^2$, as a consequence of $d(\mathbf{Z}, \mathbf{x}_1), \ldots, d(\mathbf{Z}, \mathbf{x}_k)$ being approximately Normal; and, (iii) it appraises balance not only on $\mathbf{x}_1, \ldots, \mathbf{x}_k$, but also on all linear combinations of them. Large imbalances on the linear predictor of a response variable from $\mathbf{x}_1, \ldots, \mathbf{x}_k$, for example, will make $d^2(\mathbf{z}, \mathbf{x}_1, \ldots, \mathbf{x}_k)$ large relative to its null distribution. The test is a first cousin of Hotelling's (1931) $T$-test, which treats $\mathbf{x}_1, \ldots, \mathbf{x}_k$ rather than $\mathbf{z}$ as random and is $F$-distributed, rather than $\chi^2$-distributed, under the null of equivalence between groups.

Linearity of $d(\mathbf{z}, \cdot)$ immediately establishes (iii). Arguments of Sections 3.1 and 3.2 entail that $d(\mathbf{Z}, \beta_1 \mathbf{x}_1 + \cdots + \beta_k \mathbf{x}_k)$, suitably scaled, must be asymptotically $N(0, 1)$ provided the $\mathbf{x}_i$'s are suitably regular, whatever $\beta_1, \ldots, \beta_k$ may be. It follows that the vector $[d(\mathbf{z}, \mathbf{x}_1), \ldots, d(\mathbf{z}, \mathbf{x}_k)]$ has a multivariate Normal distribution in large samples, showing (ii). Then $d^2(\mathbf{Z}; \mathbf{x}_1, \ldots, \mathbf{x}_k)$ is scalar-valued with a large-sample $\chi^2$ distribution on $\text{rank}(\text{Cov}([d(\mathbf{z}, \mathbf{x}_1), \ldots, d(\mathbf{z}, \mathbf{x}_k)]))$ degrees of freedom.

To calculate $d^2(\mathbf{z}; \mathbf{x}_1, \ldots, \mathbf{x}_k)$, one begins as if calculating each of $(d(\mathbf{z}; \mathbf{x}_i : i = 1, \ldots, k)$ separately (Section 3.4). With RITOOLS, the `xBalance` function can calculate each of these simultaneously; in this case, it optionally returns $d^2(\mathbf{z}; \mathbf{x}_1, \ldots, \mathbf{x}_k)$ and its corresponding degrees of freedom. Without this aid, the joint calculation differs from a sequence of univariate balance assessments only in requiring that covariance matrices, rather than scalars, be scaled and summed across blocks $b$, and requiring the rank and a generalized inverse of the resulting sum.

The $\chi^2$-approximation seems to work reasonably well even in small samples. Its distribution in one small simulation experiment is graphed in the right panel of Figure 1, while Table 3 summarizes its distribution in another; in both cases it tends somewhat toward conservatism. As a practical tool for the data analyst, it has the important advantage that it stably handles saturation with $x$-variables; one would not bring about a spurious rejection of the hypothesis of balance by adding to the list of $x$-variables to be tested. One certainly would decrease the test's power to detect imbalance along individual $\mathbf{x}_i$'s included among covariates tested, but that is to be expected. (An example is given in Section 6.2.) This is in important contrast with methods based on regression of $\mathbf{z}$ on $\mathbf{x}$'s; as the left panel of Figure 1 shows, natural tendencies toward overfitting inflate the Type I errors of such tests.

## 5. OPTIMIZING LOCAL POWER

This section develops and analyzes a statistical model of the *absence* of balance that is appropriate to randomization inference. Casting this model as an alternative to the null hypothesis of balance, tests based on $d$ or $d^2$ are seen to have greatest power when weights $w_b^* \propto h_b \bar{m}_b$ are used to combine differences across blocks or matched sets. Readers not seeking justification of this may prefer to skip to Section 6.

Say balance is to be assessed against a canonical model (Section 3.2) with $B$ blocks, perhaps after conditioning as in Section 3.3. What choice of weights $w_1, \ldots, w_B$ maximizes the power of the test for balance? Common results give the answer for models positing that $\mathbf{x}$ is sampled while $\mathbf{z}$ is held fixed. Kalton (1968), for instance, assumes random sampling from $2B$ superpopulations with means $\mu_{t1}, \mu_{c1}, \ldots, \mu_{tB}, \mu_{cB}$. He finds that in order to maximize power against alternatives to the effect that $\mu_{tb} \equiv \mu_{cb} + \delta$, $\delta \neq 0$, blocks' differences of means should be weighted in proportion to the inverse of the variance of those differences. With the simplifying assumption of a common variance in the $2B$ superpopulations, this leads to harmonic mean weighting, $w_b \propto h_b \bar{m}_b$. To avoid this simplification, weights might be set in proportion to reciprocals of estimated variances. But such a procedure would seem to add complexity, and to detract from the credibility of assessments of statistical significance, since the sample-to-sample fluctuation it imposes on the weighting scheme is difficult to account for at the stage of analysis (Yudkin and Moher (2001), page 347).

The randomization perspective leads to the same result, but by a cleaner route, avoiding the need to estimate or make assumptions about dispersion in superpopulations. In support of this claim, we analyze the problem of distinguishing unbiased from biased sampling of treatment assignment configurations, $\mathbf{z}$'s, from $A$, rather than differences in superpopulations from which treatment and control $x$'s are supposed to be drawn. This amounts to distinguishing constant from nonconstant $\psi_x(\cdot)$ in model (7).

Our analysis is asymptotic, assuming increasing sample size. Since any nontrivial test would have overwhelming power given a limitless stock of similarly informative observations, we mount an analysis of local power, in which the observations become less informative as sample size increases. This is modeled with $x$'s that cluster increasingly around a single value as their number increases, while bias in assignment to treatment is dictated by the same $\psi_x$. The strata may increase in number or in size, or in both, as the number of assignment units increases; it is assumed that cluster size is bounded and that the fractions of blocks allocated to treatment $n_{tb}/n_b$ are bounded away from 0 and 1. Because the observations are neither independent (due to conditioning on $\mathbf{Z}_b^t \mathbf{1}, b = 1, \ldots, B$) nor identically distributed, the asymptotic analysis pertains not to a single sequence of observations but to a sequence of experimental populations $\nu = 1, 2, \ldots$ containing increasing numbers of observations.

Conditions A1–A4, stated in the Appendix, entail certain convergences of weights and variances, at least along subsequences $\{\nu_i\}$ of populations. Specifically, with $\{\nu\}$ narrowed to such a subsequence there are positive constants $K, s_{0x}, s_{wx}$ and $v_{wx}$ such that as $\nu \to \infty$,

$$(10) \quad \begin{aligned} n_\nu^{-1} \sum_b \bar{m}_{\nu b} h_{\nu tb} &\to K \quad \text{and} \\ n_\nu \sum_b w_{\nu b}^* \frac{s^2(\mathbf{x}_{\nu b})}{\bar{m}_{\nu b}} &\to s_{0x}^2; \end{aligned}$$

$$(11) \quad \begin{aligned} n_\nu \sum_b w_{\nu b} \frac{s^2(\mathbf{x}_{\nu b})}{\bar{m}_{\nu b}} &\to s_{wx}^2 \quad \text{and} \\ n_\nu^2 \operatorname{Var}_P(d(\mathbf{Z}_\nu, \mathbf{x}_\nu)) &\to v_{wx}^2, \end{aligned}$$

where $d(\mathbf{Z}_\nu, \mathbf{x}_\nu)$ in (11) is understood in the sense of (12).

PROPOSITION 5.1. *Let*

$$(12) \quad \begin{aligned} &d(\mathbf{Z}_\nu, \mathbf{x}_\nu) \\ &= \sum_b w_{\nu b} \left[ \frac{\mathbf{Z}_{\nu b}^t \mathbf{x}_{\nu b}}{\bar{m}_{\nu b} n_{\nu tb}} - \frac{(\mathbf{1} - \mathbf{Z}_{\nu b})^t \mathbf{x}_{\nu b}}{\bar{m}_{\nu b}(n_{\nu b} - n_{\nu tb})} \right]. \end{aligned}$$

*Assume conditions* A1–A4, *write $P$ and $Q$ for distributions of $\mathbf{Z}_\nu$ under, respectively, the null of unbiased assignment and the alternative of bias according to (7) with nonconstant $\psi$, and let $s_{wx}, v_{wx}$ be*

as in (11). Then

$$\mathbf{P}_Q(d(\mathbf{Z}_\nu, \mathbf{x}_\nu) > z^* \operatorname{Var}_P(d(\mathbf{Z}_\nu, \mathbf{x}_\nu))^{1/2})$$
$$\to 1 - \Phi\left(z^* - \beta \frac{s_{wx}^2}{v_{wx}}\right), \tag{13}$$

where $\beta$ is the derivative of $\psi$ at $c$ (as defined in condition A3).

For a proof, see the Appendix.

Compare (13) to

$$\mathbf{P}_P(d(\mathbf{Z}_\nu, \mathbf{x}_\nu) > z^* \operatorname{Var}_P(d(\mathbf{Z}_\nu, \mathbf{x}_\nu))^{1/2}) \to 1 - \Phi(z^*),$$

a statement of the asymptotic normality of $d(\mathbf{Z}_\nu, \mathbf{x}_\nu)$ under the null hypothesis: in the limit, the amount by which power exceeds size increases with the ratio $s_{wx}^2/v_{wx}$. Specifically, if the acceptance region is limited from above at $z_u \operatorname{Var}_P(d(\mathbf{Z}_\nu, \mathbf{x}_\nu))^{1/2}$, $z_u > 0$, then power against alternatives with $\beta > 0$ is optimized by calibrating the stratum weights $(w_{\nu b})$ so as to maximize $\operatorname{Var}_P(d(\mathbf{Z}_\nu, \mathbf{x}))^{-1/2}(\sum_b w_{\nu b} s^2(\mathbf{x}_{\nu b})/\bar{m}_{\nu b})$, the limit of which is $s_{wx}^2/v_{wx}$. (If the acceptance region has a finite lower limit, then a symmetric argument yields that the same calibration maximizes power against alternatives with $\beta < 0$.) To effect this calibration, for the moment fix $\nu$ and write $s_b^2$ for $s^2(\mathbf{x}_{\nu b})/\bar{m}_{\nu b}$, $b = 1, \ldots, B$. Recall that $h_b = [n_{tb}(1 - n_{tb}/n_b)]$ (Section 3.2). Then

$$\frac{(\sum_b w_{\nu b} s^2(\mathbf{x}_{\nu b})/\bar{m}_{\nu b})}{\operatorname{Var}_P(d(\mathbf{Z}_\nu, \mathbf{x}))^{1/2}}$$

$$= \frac{(\sum_b w_b s_b^2)}{(\sum_b w_b^2 [h_b \bar{m}_b]^{-1} s_b^2)^{1/2}}$$

$$= \frac{((w_b [h_b \bar{m}_b]^{-1/2} s_b) : b \leq B)^t}{\|((w_b [h_b \bar{m}_b]^{-1/2} s_b) : b \leq B)\|_2} \tag{14}$$

$$\cdot ([h_b \bar{m}_b]^{1/2} s_b : b \leq B),$$

where $\|\cdot\|_2$ is the Euclidean norm, $\|\mathbf{x}\|_2 = (\sum_i x_i^2)^{1/2}$. Selecting $(w_b : b = 1, \ldots, B)$ so as to maximize this expression amounts to maximizing the correlation between $B$-dimensional vectors $(w_b [h_b \bar{m}_b]^{-1/2} s_b : b = 1, \ldots, B)$ and $([h_b \bar{m}_b]^{1/2} s_b : b = 1, \ldots, B)$, which is achieved by setting $w_b \propto h_b \bar{m}_b$—that is, by $w_{\nu b} = w_{\nu b}^*$.

## 6. APPLICATIONS TO STUDY DESIGN AND ANALYSIS

### 6.1 Whether to Stratify, and Which Stratification is Best

Randomization within well-chosen blocks may lead to imbalances on baseline measures of smaller absolute magnitude than unrestricted randomization, and smaller baseline imbalances are preferable for various reasons. Raab and Butcher (2001) sought to avoid imbalances large enough to create noticeable discrepancies between treatment effects estimated with and without covariance adjustment. Such differences might be troubling to the policymakers who were a central audience for their study, even if they fell within estimated standard errors. Yudkin and Moher (2001) worry that designs in which sizable imbalances are possible may sacrifice power.

To head off these problems, Yudkin and Moher's ASSIST team elected to randomize clinics within three blocks, consisting of 6, 9 and 6 clinics, rather than to randomly assign treatment to 14 of 21 clinics outright. It remained to be decided which baseline variable to block on. They report deciding against blocking on clinic size after finding only weak correlations between clinics' sizes and baseline rates of adequate heart disease assessments; they feared that privileging size in the formation of blocks could have "resulted in imbalance in the main prognostic factor" (Yudkin and Moher (2001), page 345). While these correlations are certainly reasonable to consider, it might have been more direct to compare candidate blocking schemes on the basis of the variance in $d(\mathbf{Z}, \cdot)$'s they would entail, preferring those schemes that offer lesser mean-square imbalances on key prognostic variables.

Table 4 offers such a comparison. It emerges that, despite the weak relationship between clinic size and baseline rate of adequate assessment, blocking on size balances the rate of adequate assessment quite well, nearly as well as does blocking on the rate itself. Meanwhile, to balance other baseline variables, rates of treatment with various drugs that at followup would be measured as secondary outcomes, it is much better to block on size. [Lewsey (2004) discusses size blocking in some detail.] Perhaps the investigators were too quick to reject this option. In any case, the comparison of $\operatorname{Var}(d(\mathbf{Z}, \mathbf{x}))$, from (6), for various blocking schemes and covariates, $\mathbf{x}$, would more directly have informed their decision.

### 6.2 Whether to Poststratify, and Whether a Given Poststratification Suffices

Comparative studies typically present a small number of covariates that *must* be balanced in order for the study to be convincing, along with a longer list of variables on which balance would be advantageous.

TABLE 4
*Standard deviations of $d(\mathbf{Z}, \mathbf{x})$ under various stratification schemes, expressed as fractions of an s.d. of $\mathbf{x}/\bar{m}$*

| | Baseline variable | | | |
|---|---|---|---|---|
| | Adequate | Treatment with | | |
| Stratification | assessment | aspirin | hypotensives | lipid-reducers |
| None | 0.46 | 0.46 | 0.46 | 0.46 |
| By rate of adequate assessment | 0.31 | 0.42 | 0.43 | 0.36 |
| By clinic size | 0.33 | 0.24 | 0.24 | 0.31 |

Both stratification schemes offer distinctly better expected balance than no stratification at all, and stratification on clinic size seems preferable to stratification on clinics' baseline rates of adequate assessment.

In the ASSIST a trial, the short list consists of baseline measures on variables to be used as outcomes; in the Vote'98 experiment, it comprises a "baseline" measure of the outcome, voting in the previous election, along with party membership and demographic data that predict voting. Were treatment subjects appreciably older, and so perhaps more likely to vote (Highton and Wolfinger, 2001) than controls, or were they more likely to have voted in past elections, then one would suspect appreciable positive error in unadjusted estimates of the treatment effect—even in the presence of randomization, which controls such errors *most* of the time but not *all* of the time.

Even if discovered only after treatments have been applied, such imbalances can be remedied by poststratification: if treatments are on the whole older than controls, for example, then compare older treatments only to older controls, and also compare younger subjects only among themselves. There is the possibility that one could introduce imbalances on other variables by subclassifying on age; to assess this, one might apply $d^2(\mathbf{z}; \mathbf{x}_1, \ldots, \mathbf{x}_k)$, where $\mathbf{x}_1, \ldots, \mathbf{x}_k$ make up the short list, to the poststratified design. Should subclassifying only on age fail to sufficiently reduce $d^2(\mathbf{z}; \mathbf{x}_1, \ldots, \mathbf{x}_k)$, or should there be a more complex pattern of misalignment to begin with, propensity-score methods are a reliable alternative(Rosenbaum and Rubin (1984), and Hillet at al., 2000). Indeed, with the option of propensity score subclassification, there is little reason to restrict one's attention entirely to the short list; one can reasonably hope to relieve gross imbalances on any of a longer list of covariates, as well as smaller imbalances on the most important ones.

Perhaps with this in mind, Imai (2005) suggests checking the Vote'98 data for imbalance twice, once focusing on short-list variables and a second time considering also second-order interactions of them.

As discussed by Arceneaux et al. (2004), and as the discussion of Section 2 would predict, his logistic-regression based check gives misleading results. Despite this technical impediment, however, the spirit of the suggestion is sound; one might hope the check based on $d^2$ would perform more reliably. In fact it does: in $10^6$ simulated reassignments of telephone GOTV, the $d^2$ statistic combining imbalances on all first- and second-order interactions of $x$-variables exceeded nominal 0.001, 0.01, 0.05 and 0.10 levels of the $\chi^2(363)$ distribution in 0.09%, 0.9%, 4.8%, and 9.7% of trials, respectively. The treatment assignment actually used gives, for the long list, $d^2 = 360.6$, with theoretical and simulation $p$-values 0.526 and 0.527, respectively, and for the short list, $d^2 = 26.6$ on 38 d.f.'s, with $p$-values 0.918 and 0.918, respectively; it is well balanced.

## 7. SUMMARY AND DISCUSSION

Clinical trials methodologists note, with some alarm, how few cluster randomized trials explicitly make note of cluster-level assignment and account for it in the analysis (Divine et al. (1992); MacLennan et al. (2003); Isaakidis and Ioannidis (2003)). We have seen the need for such an accounting even when it seems least necessary, with clusters that are small, uniform in size, and numerous. We have also seen that one natural model-based test for balance along covariates, the test based on logistic regression, is prone to spuriously indicate lack of balance when there are too many covariates relative to observations, and that this condition obtains for surprisingly large ratios of observations to the number of covariates.

Cluster-level randomization is said to confront investigators with "a bewildering array of possible approaches to the data analysis" (Donner and Klar, 1994). Randomization inference presents a less cluttered field of options, and has the additional advan-tages of adaptation specifically to comparative studies and of being nonparametric. With appropriate attention to the form of the test statistic, it is quite possible in the randomization framework to respect the study's design while training attention on differences among individuals. This aim also suggests conditioning strategies appropriate to the problem of assessing covariate balance. The result is a class of test statistics that one can expediently appraise using Normal approximations which are quite accurate in small and moderate samples. The tests gauge balance on a single covariate or on a set of covariates jointly; in the latter case, they also implicitly assess imbalance on linear combinations of the covariates, including projections of the response variable into covariate space. Our analysis of a model of biased assignment suggests values for tuning parameters that completely specify, indeed simplify, the form of the resulting nonparametric tests, ending with a simple prescription that is suitable for general use: assess balance along individual covariates $\mathbf{x}$ with the differences $d(\mathbf{z}, \mathbf{x})$ between treatment and control groups' adjusted means, using weights $w_b \propto \bar{m}_b h_b$ to combine across blocks as in (5); then mount an overall test by referring the combined baseline difference $d^2(\mathbf{z}; \mathbf{x}_1, \ldots, \mathbf{x}_k)$, as in (9), to the appropriate $\chi^2$-distribution.

As an omnibus measure of balance, the combined baseline difference statistic $d^2(\mathbf{z}; \mathbf{x}_1, \ldots, \mathbf{x}_k)$ is similar in form and spirit to a statistic suggested by Raab and Butcher (2001), namely a weighted sum of squares of differences of means of cluster means: $\alpha_1 d(\mathbf{z}, \mathbf{x}_1/\mathbf{m})^2 + \cdots + \alpha_k d(\mathbf{z}, \mathbf{x}_k/\mathbf{m})^2$, where $\alpha_1, \ldots, \alpha_k \geq 0$ sum to 1. The ability of the statistician to decide the relative weightings $\alpha$ of the variables might in some contexts be an advantage, but in others it may be burdensome. In all cases it lends some arbitrariness to the criterion. Also, the criterion directly measures only imbalances in $\mathbf{x}_1, \ldots, \mathbf{x}_k$. In contrast, $d^2(\mathbf{z}; \mathbf{x}_1, \ldots, \mathbf{x}_k)$ measures imbalances in linear combinations of $\mathbf{x}_1, \ldots, \mathbf{x}_k$ as much as in these variables themselves, lets the data drive the weighting scheme, upweighting discrepancies along variables with less variation in general, and has the advantage of easy calibration against $\chi^2$ tables.

Altman (1985), Begg (1990) and Senn (1994) criticize the use of balance tests to decide which covariates to adjust for in the outcome analysis of a clinical trial, arguing that these judgments should rather be made on the basis of the prognostic value of the covariate. These criticisms are sometimes taken to support the stronger conclusion that balance tests are inappropriate for any purpose. The criticisms do not, however, speak against the use of balance tests to detect problems of implementation, nor do they preclude a possible role for assessments of balance in the interpretation of study results (Begg, 1990). Indeed, the CONSORT statement on reporting in clinical trials (Begg et al., 1996) recommends that reports include assessments of balance on variables of possible prognostic value.

Section 5 established optimality of tests based on $d$ and $d^2$ within one class of balance criteria and under certain conditions, but in some settings other statistics may be better equipped to reveal biased assignment. For instance, in some clinical trials that enroll patients sequentially and at the discretion of their physicians it is possible for the physician to guess or infer the treatment to which a potential patient would be assigned; the methods of Berger and Exner (1999) and Berger (2005) model patterns of assignment that would occur if physicians were using this foreknowledge to the advantage of one assignment arm or the other, and may have greater power in such situations.

When subclassifying or matching on the propensity score, systematic appraisals of balance are needed to check and tune the propensity adjustment (Rosenbaum and Rubin, 1984). An *exact* propensity stratification would make an observational study as well-balanced as if its treatment conditions had been assigned randomly within the strata, but the inevitably more crude propensity stratifications that are available in practice may yield less balance. Balance tests based on $d^2$ are particularly well-suited to adjudicate the success or failure of a given *inexact* propensity model and stratification procedure. In contrast with the case of randomized assignment, propensity adjustment inevitably leaves at least some within-stratum variation in probabilities of assignment to treatment, making it certain that the hypothesis of unbiased allocation is *false*, at least in detail (Hansen, 2008). One hopes, however, that the bias is sufficiently small so as not to imbalance covariates discernibly more than random assignment would be expected to have done, and this is precisely the question that $d^2$ addresses. With its focus on the randomization distribution, it avoids modeling treatment and control observations as having been sampled from respective superpopulations, an undesirable feature of many other balance tests (Imai et al., 2008). Another advantage of $d$ and $d^2$ for observational studies is that they apply without modification to matched data, by treating the matched sets as strata; likelihood ratio tests of logistic regression models, by contrast, are not consistent when used in this way (Agresti (2002), Section 6.3.4).

## APPENDIX: PROOF OF PROPOSITION 5.1

Let there be constants $\{x_{\nu b i}\}$, $\{m_{\nu b i}\}$, and random indicator variables $\{Z_{\nu b i}\}$, arranged in triangular arrays the $\nu$th rows of which contain $n_\nu$ entries, $\mathbf{x}_\nu^t = (\mathbf{x}_{\nu 1}^t, \ldots, \mathbf{x}_{\nu B_\nu}^t)$, $\mathbf{m}_\nu^t = (\mathbf{m}_{\nu 1}^t, \ldots, \mathbf{m}_{\nu B_\nu}^t)$ and $\mathbf{Z}_\nu^t = (\mathbf{Z}_{\nu 1}^t, \ldots, \mathbf{Z}_{\nu B_\nu}^t)$, respectively, where $\mathbf{x}_{\nu b} = (x_{\nu b 1}, \ldots, x_{\nu b n_{\nu b}})^t$, $\mathbf{m}_{\nu b} = (m_{\nu b 1}, \ldots, m_{\nu b n_{\nu b}})^t$ and $\mathbf{Z}_{\nu b} = (Z_{\nu b 1}, \ldots, Z_{\nu b n_{\nu b}})^t$, for some whole numbers $B_\nu$ and $n_{\nu 1}, \ldots, n_{\nu B_\nu}$. Within a given row $\nu$, $\mathbf{x}_{\nu b}, \mathbf{m}_{\nu b}$, and $\mathbf{Z}_{\nu b}$ describe cluster totals on a variable $x$, cluster sizes (averaging to $\bar{m}_{\nu b}$) and treatment assignments within block $b$, any $b \leq B_\nu$. Suppose $1 \leq n_{\nu t b} < n_{\nu b}$ for all $\nu, b$, and assume of the random variables $\mathbf{Z}_{\nu b}$ that with probability 1, $\mathbf{Z}_{\nu b}^t \mathbf{1} = n_{\nu t b}$ for each $\nu$ and $b \leq B_\nu$; say vectors $\mathbf{z}_\nu$ that lack this property are *excluded*. The null hypothesis asserts that for each $\nu$ and block $b \leq B_\nu$, $\mathbf{P}(Z_{\nu b i} = 1)$ is the same for all indices $i$. Alternately put, its probability density $P(\mathbf{z}_\nu)$ vanishes for excluded $\mathbf{z}_\nu$ and otherwise is proportional to (7) with $\psi_x \equiv 0$. For alternatives $Q$ to this null, define (for nonexcluded $\mathbf{z}_\nu$) a likelihood proportional to (7), with bias function $\psi_x(\cdot)$ the same for all $\nu$. Assume of this sequence of models that:

A1 $\{m_{\nu b i}\}$ is uniformly bounded, and $\{n_{\nu t b}/n_{\nu b}\}$ is uniformly bounded away from 0 and 1;
A2 weights $w_{\nu b}$ have the property that $w_{\nu b}/w_{\nu b}^*$ is uniformly bounded away from 0 and $\infty$, where $w_{\nu b}^* \propto \bar{m}_{\nu b} n_{\nu t b}(1 - n_{\nu t b}/n_{\nu b})$ and $\sum_b w_{\nu b}^* = 1$;
A3 for some $c$, $\sup_{b,i}|x_{\nu b i} - c| \downarrow 0$ and $\sum_{b \leq B_\nu} \sum_i (x_{\nu b i} - c)^2$ is $O(1)$ as $\nu \to \infty$;
A4 $\psi_x$ is differentiable at $c$, where $c$ is the constant referred to in A3.

Condition A1 has the side-effect of limiting the divergence of $w_{\nu b}^*$ and other common weighting schemes; should weights $w_{\nu b}$ be proportional to the number of subjects in a block, the number of treatment group subjects in a block, or the total of controls by block, then by condition A1, $w_{\nu b}^*/w_{\nu b}$ will be universally bounded away from 0 and $\infty$. In other words, given A1, condition A2 is not restrictive. Condition A1 also ensures that $\sum_b h_{\nu b} \bar{m}_{\nu b}$ is $O(n_\nu)$. Condition A3 ensures tightening dispersion of $x$'s around $c$. In particular, combined with A1, condition A3 entails $\sum_b w_{\nu b}^* s_{\nu b}^2(\mathbf{x}_{\nu b})/\bar{m}_{\nu b}$ is $O(n_\nu^{-1})$, or that with weighting by either $w_\nu^*$ or $w_\nu$, the weighted average of block mean differences $d(\mathbf{Z}_\nu, \mathbf{x}_\nu)$ has variance of order $O(n_\nu^{-2})$—see Proposition 3.1 and Corollary 3.1.

We establish Proposition 5.1 using principles of contiguity (Le Cam (1960); Hájek and Šidák (1967)), which describe the limiting $Q$-distribution of a test statistic $t(\mathbf{Z})$ in terms of the limiting joint distribution, *under* $P$, of $(t(\mathbf{Z}), \log \frac{dQ}{dP}(\mathbf{Z}))$. A technical lemma, Lemma A.1, is needed, after which contiguity results are invoked to establish Lemma A.2 (from which the proposition is immediate).

LEMMA A.1. *Under the hypotheses of Proposition 5.1,*

$$\log \frac{dQ}{dP}(\mathbf{Z}) \stackrel{P}{\Rightarrow} N\left(-\frac{1}{2}\beta^2 K s_{0x}^2, \beta^2 K s_{0x}^2\right)$$

*(where "$\stackrel{P}{\Rightarrow}$" denotes convergence in distribution under $P$).*

LEMMA A.2. *Under the hypotheses of Proposition 5.1,*

$$\frac{d(\mathbf{Z}_\nu, \mathbf{x}_\nu)}{\sqrt{\operatorname{Var}_P(d(\mathbf{Z}_\nu, \mathbf{x}_\nu))}} \stackrel{Q}{\Rightarrow} N\left(\beta \frac{s_{wx}^2}{v_{wx}}, 1\right).$$

### A.1 Proof of Lemma A.1

Without loss of generality, the $c$ named in conditions A3 and A4 is 0. Then one has $\psi_x(x) = \psi_x'(0)x + o(|x|) = \beta x + x e(x)$, where, because of condition A3, $\max_{b,i}|e(x_{\nu b i})| \downarrow 0$ as $\nu \uparrow \infty$. Since $Q$ is defined by (7) and $P$ is defined by (7) without the $\psi_x$ term, one can write

$$\begin{aligned}
\log &\frac{dQ}{dP}(\mathbf{Z}_{\nu b}) \\
&= \beta \sum_b \mathbf{Z}_{\nu b}^t (\mathbf{x}_{\nu b} - \bar{x}_{\nu b}) \\
(A.1) \quad &\quad + \sum_b \mathbf{Z}_{\nu b}^t (\mathbf{x}_{\nu b} e(\mathbf{x}_{\nu b}) - \overline{x_{\nu b} e(x_{\nu b})}) \\
&\quad + \kappa_{\nu P} - \kappa_{\nu Q} \\
&=: X_\nu + Y_\nu - (\kappa_{\nu Q} - \kappa_{\nu P}),
\end{aligned}$$

for appropriate constants $\kappa_{\nu P}, \kappa_{\nu Q}$.

By calculations similar to those justifying Proposition 3.1, $\operatorname{Var}_P(X_\nu) = \beta^2 \sum_b h_{\nu b} s^2(\mathbf{x}_{\nu b})$. By condition A3 and (10), this variance approaches $\beta^2 K s_{0x}^2$. By the discussion following (3), $\operatorname{Var}_P(Y_\nu) = \sum_b h_{\nu b} \times s^2(\mathbf{x}_{\nu b} e(\mathbf{x}_{\nu b}))$. By condition A3, this is $O(e_\nu^2)$ as $\nu \uparrow$ $\infty$, where $e_\nu := \sup_{b,i} |e(x_{\nu b i})|$. By condition A3 and A4, of course, $e_\nu \downarrow 0$ as $\nu \uparrow \infty$; thus $\mathrm{Var}_P(Y_\nu) \downarrow 0$ as $\nu \uparrow \infty$. Since (as we have seen) $\mathrm{Var}_P(X_\nu)$ is $O(1)$, it follows also that $\mathrm{Cov}_P(X_\nu, Y_\nu) = O(e_\nu)$, and overall $\mathrm{Var}_P(X_\nu + Y_\nu) \to \beta^2 K s_{0x}^2$ as $\nu \uparrow \infty$.

Clearly both $X_\nu$ and $Y_\nu$ have expectation 0, under $P$. Since the random term $X_\nu + Y_\nu$ is, as in Section 3.2, a sum of totals of simple random samples, its limiting law (under $P$) must be $N(0, \beta^2 K s_{0x}^2)$.

It remains to be shown that $\kappa_{\nu Q} - \kappa_{\nu P} \to \frac{1}{2}\beta^2 K s_{0x}^2$. Since $\mathbf{E}_P((dQ/dP)(\mathbf{Z})) = 1$, $\exp\{\kappa_{\nu Q} - \kappa_{\nu P}\} = \mathbf{E}_P(e^{X_\nu + Y_\nu})$. From what was just shown it follows immediately that $e^{X_\nu + Y_\nu} \stackrel{P}{\Rightarrow} e^{N(0, \beta^2 K s_{0x}^2)}$, the expectation of which equals the moment generating function of the standard Normal distribution evaluated at $\beta K^2 s_{0x}^2$, or $\exp\{\frac{1}{2}\beta^2 s_{0x}^2\}$. So the conclusion follows if we can establish that $\mathbf{E}_P(e^{X_\nu + Y_\nu})$ converges to $\mathbf{E}(e^{N(0, \beta^2 K s_{0x}^2)})$. This would follow from uniform integrability of the random variables $e^{X_\nu + Y_\nu}$, which would follow in turn from $\sup_n \mathbf{E}_P(e^{(1+\varepsilon)(X_\nu + Y_\nu)}) < \infty$, any $\varepsilon > 0$.

The rest of the argument verifies this by establishing the technical condition that $\limsup_\nu \mathbf{E}_P(\exp\{\sqrt{2}(X_\nu + Y_\nu)\}) < \infty$. We make use of a theorem of Hoeffding (1963), to the effect that the expectation of a convex continuous function of a sum of a simple random sample is bounded above by the expectation of the same function of a similarly sized with-replacement sample from the same population, and of the fact from calculus that if for a triangular array $\{c_{ij}\}$ of nonnegative numbers, $\max_j c_{ij} \downarrow 0$ while $\sum_j c_{ij} \to \lambda$, then $\prod_j (1 + c_{ij}) \to e^\lambda$. Write $m_{\nu b}(t)$ for the moment generating function of $\mathbf{Z}_{\nu b}^t (\psi_x(\mathbf{x}_{\nu b}) - \overline{\psi_x(x)}_{\nu b})$, so that $\mathbf{E}_P(e^{t(X_\nu + Y_\nu)}) = \prod_b m_{\nu b}(t)$. Under $P$, $\mathbf{Z}_{\nu b}^t(\psi_x(\mathbf{x}_{\nu b}) - \overline{\psi_x(x)}_{\nu b})$ is the sum of a simple random sample of size $n_{\nu t b}$, and by Hoeffding's theorem $m_{\nu b}(t) \leq (\tilde{m}_{\nu b}(t))^{n_{\nu t b}}$, where $\tilde{m}_{\nu b}(t)$ is the moment generating function of a single draw, $D_{\nu b}$, from $\{\psi_x(\mathbf{x}_{\nu b i}) - \overline{\psi_x(x)}_{\nu b i} : i \leq n_{\nu b}\}$. By Taylor approximation, for each $\nu$ and $b$, $\tilde{m}_{\nu b}(\sqrt{2}) = 1 + \mathbf{E}_p(D_{\nu b}^2 \exp\{t_{\nu b}^* D_{\nu b}\})$, some $t_{\nu b}^* \in [0, \sqrt{2}]$. We now need to show that $\max_b \mathbf{E}_p(D_{\nu b}^2 \exp\{t_{\nu b}^* D_{\nu b}\}) \downarrow 0$ and $\sum_b n_{\nu t b} \mathbf{E}_p(D_{\nu b}^2 \exp\{t_{\nu b}^* D_{\nu b}\})$ is $O(1)$. By condition A3, as $\nu$ increases $D_{\nu b}^2 \times \exp\{t_{\nu b}^* D_{\nu b}\}$ is deterministically bounded by constants tending to 0, entailing $\max_b \mathbf{E}_p(D_{\nu b}^2 \exp\{t_{\nu b}^* D_{\nu b}\}) \downarrow 0$. $\exp\{t_{\nu b}^* D_{\nu b}\}$ also declines to 0 deterministically, so that the sum of $n_{\nu t b} \mathbf{E}_p(D_{\nu b}^2 \exp\{t_{\nu b}^* D_{\nu b}\})$ is $O(1)$ if $\sum_b n_{\nu t b} \mathbf{E}_p(D_{\nu b}^2) = \sum_b n_{\nu t b} \sigma^2(\psi_x(\mathbf{x}_{\nu b}))$ is. Now $\sum_b n_{\nu t b} \sigma^2(\psi_x(\mathbf{x}_{\nu b})) = \sum_b n_{\nu t b} \beta^2 \sigma^2(\mathbf{x}_{\nu b}) + \sum_b n_{\nu t b} \sigma^2(\psi_x(\mathbf{x}_{\nu b}) - \beta \mathbf{x}_{\nu b}) + 2\sum_b n_{\nu t b} \beta \sigma(\mathbf{x}_{\nu b}, \psi_x(\mathbf{x}_{\nu b}) - \beta \mathbf{x}_{\nu b})$. Invoking condition A3, the first of these three sums may be seen to be $O(1)$, and the latter two $O(e_\nu^2)$ and $O(e_\nu)$, respectively, as $\nu \uparrow \infty$. It follows that $\prod_b (\tilde{m}_{\nu b}(\sqrt{2}))^{n_{\nu t b}}$, and hence $\prod_b m_{\nu b}(\sqrt{2})$, are $O(1)$, confirming that $\{e^{X_\nu + Y_\nu} : l = 1, \ldots\}$ is uniformly integrable.

### A.2 Proof of Lemma A.2

Write $T_\nu := \mathrm{Var}_P(d(\mathbf{Z}_\nu, \mathbf{x}_\nu))^{-1/2} d(\mathbf{Z}_\nu, \mathbf{x}_\nu)$. By arguments of Section 3.2, $T_\nu \stackrel{P}{\Rightarrow} N(0,1)$. Combining this with Lemma A.1, one has that

$$\left(T_\nu, \log \frac{dQ}{dP}(\mathbf{Z}_\nu)\right) \stackrel{P}{\Rightarrow} N\left[(0, -\sigma^2/2), \begin{pmatrix} 1 & r \\ r & \sigma^2 \end{pmatrix}\right],$$

for some as yet to be determined $r$. This establishes the premise of Le Cam's Third Lemma (Le Cam (1960); Hájek and Šidák (1967)), the conclusion of which is that the limit law *under* $Q$ of the random variable $T_\nu$ is $N(r, 1)$. We now calculate $r$.

Using the notation of (A.1), $\mathrm{Cov}(T_\nu, \log \frac{dQ}{dP}(\mathbf{Z}_\nu)) = \mathrm{Cov}(T_\nu, X_\nu) + \mathrm{Cov}(T_\nu, Y_\nu)$. Now $|\mathrm{Cov}(T_\nu, Y_\nu)| \leq (\mathrm{Var}(T_\nu)\mathrm{Var}(Y_\nu))^{1/2} = \mathrm{Var}(Y_\nu)^{1/2}$, which was shown in the proof of Lemma A.1 to decline to 0 as $\nu$ increases. Considering only nonexcluded treatment assignments $\mathbf{Z}_\nu$,

$\mathrm{Cov}_P(T_n, X_\nu)$
$= V^{-1/2} \mathrm{Cov}_P\left(\sum_{b=1}^B \frac{w_{\nu b}}{h_{\nu b} \bar{m}_{\nu b}} \mathbf{Z}_{\nu b}^t \mathbf{x}_{\nu b}, \sum_b \beta \mathbf{Z}_{\nu b}^t \mathbf{x}_{\nu b}\right)$
$= \beta V^{-1/2} \sum_b \frac{w_b}{h_{\nu b} \bar{m}_{\nu b}} \mathrm{Var}_P(\mathbf{Z}_{\nu b}^t \mathbf{x}_{\nu b})$
$= \beta V^{-1/2} \sum_b w_b s^2(\mathbf{x}_{\nu b})/\bar{m}_{\nu b},$

writing $V$ for $\mathrm{Var}_P(d(\mathbf{Z}_\nu, \mathbf{x}_\nu))$, invoking independence of $\mathbf{Z}_b$ and $\mathbf{Z}_{b'}$, $b \neq b'$, and evaluating $\mathrm{Var}_P(\mathbf{Z}_b^t \mathbf{x}_b)$ in the same manner as led to Proposition 3.1. According to (11), then, $\mathrm{Cov}_P(T_\nu, X_\nu) \to \beta \frac{s_{wx}^2}{v_{wx}}$. It follows that $r = \beta \frac{s_{wx}^2}{v_{wx}}$.

### ACKNOWLEDGMENTS

The authors thank Vance Berger, Alan Gerber, Donald Green, Kosuke Imai, Yevgeniya Kleyman, Gary King, Nancy Reid and Paul Rosenbaum for helpful discussions. Portions of this work were presented at the 2005 meetings of the Political Methodology Society and in seminars at the Department of Medicine, Case Western Reserve University, and at the Department of Biostatistics, Yale University; the authors are grateful for comments received at these venues. They also thank an anonymous reviewer, an associate editor and the executive editor, Edward George, for comments and suggestions that much improved the paper. Research support came from NICHD/NIH (#P01 HD045753-02) and the Robert Wood Johnson Scholars in Health Policy Research Program.